\documentclass[manuscript,screen]{acmart}
\AtBeginDocument{%
  }

\setcopyright{acmlicensed}
\copyrightyear{2026}
\acmYear{2026}
\acmDOI{XXXXXXX.XXXXXXX}
\acmConference[Conference acronym 'XX]{}{June 03--05,
  2026}{Woodstock, NY}
  
\acmISBN{978-1-4503-XXXX-X/2026/06}

\usepackage{booktabs}
\usepackage{multirow}
\usepackage{graphicx}
\usepackage{makecell}
\usepackage{algorithm}
\usepackage{algorithmic}
\usepackage{bm}
\usepackage{caption}
\mathchardef\mhyphen="2D

\newcommand{\ourname}{ANR-DiffRec}

\definecolor{chestnut}{cmyk}{0, 0.7808, 0.4429, 0.1412}

\begin{document}

\title{Adaptive Item-based Collaborative Structures via Noise Rescheduling in  Diffusion for Generative Recommendation}

\author{Jiaqi Wang}
\email{wangjq@tongji.edu.cn}
\authornote{Both authors contributed equally to this research.}
\orcid{0009-0008-7293-3156}
\affiliation{%
  \institution{Tongji University}
  \city{Shanghai}
  \country{China}}
\author{Tianying Liu}
\email{tianying_liu@outlook.com}
\authornotemark[1]
\orcid{0000-0003-3155-7832}
  
\author{Heng Chang}
\email{changh.heng@gmail.com}
\authornote{Corresponding authors.}
\orcid{0000-0002-4978-8041}  
\affiliation{%
  \institution{Huawei Technologies Co., Ltd.}
  \city{Shanghai}
  \country{China}}
  
\author{Jihong Guan}
\email{jhguan@tongji.edu.cn}
\authornotemark[2]
\orcid{0000-0003-2313-7635}

\author{Wengen Li}
\email{lwengen@tongji.edu.cn}
\orcid{0000-0002-8768-6740}
\affiliation{%
  \institution{Tongji University}
  \city{Shanghai}
  \country{China}}
  
\author{Shuigeng Zhou}
\email{sgzhou@fudan.edu.cn}
\orcid{0000-0002-1949-2768}
\affiliation{%
  \institution{Fudan University}
  \city{Shanghai}
  \country{China}}

\renewcommand{\shortauthors}{Wang et al.}

\begin{abstract}
Discrete Diffusion Models (DDMs) have recently been introduced to recommendation systems, modeling user history as a token generation process via iterative denoising. However, while effective at capturing user-level sequential patterns, these methods often fail to explicitly integrate item-based collaborative filtering information, a critical component for accurate recommendation. This deficiency manifests in two key aspects: (1) the item representation is often semantic-focused, lacking collaborative priors for diffusion training; and (2) the denoising process employs a uniform noise schedule, treating all tokens indiscriminately and ignoring item-level adaptive structural dependencies. To bridge this gap, we propose \ourname, a unified framework designed to encode item-based collaborative structures into discrete diffusion for generative recommendation. First, we explicitly incorporate an item co-occurrence matrix to guide semantic ID generation, providing a structured collaborative prior for discrete diffusion training. Second, we introduce an item-based adaptive noise rescheduling mechanism that dynamically adjusts denoising weights according to both local contextual recoverability and behavior-aware item dependencies. Specifically, the proposed strategy jointly models intra-item structural context and inter-item collaborative signals, enabling structure-aware denoising during diffusion training. Extensive experiments on multiple benchmarks demonstrate that our method consistently outperforms state-of-the-art generative recommendation models. Code: https://github.com/CalmaQi/ANR-DiffRec.
\end{abstract}



\maketitle

\section{Introduction}
\begin{figure}
    \centering
    \includegraphics[width=0.9\linewidth]{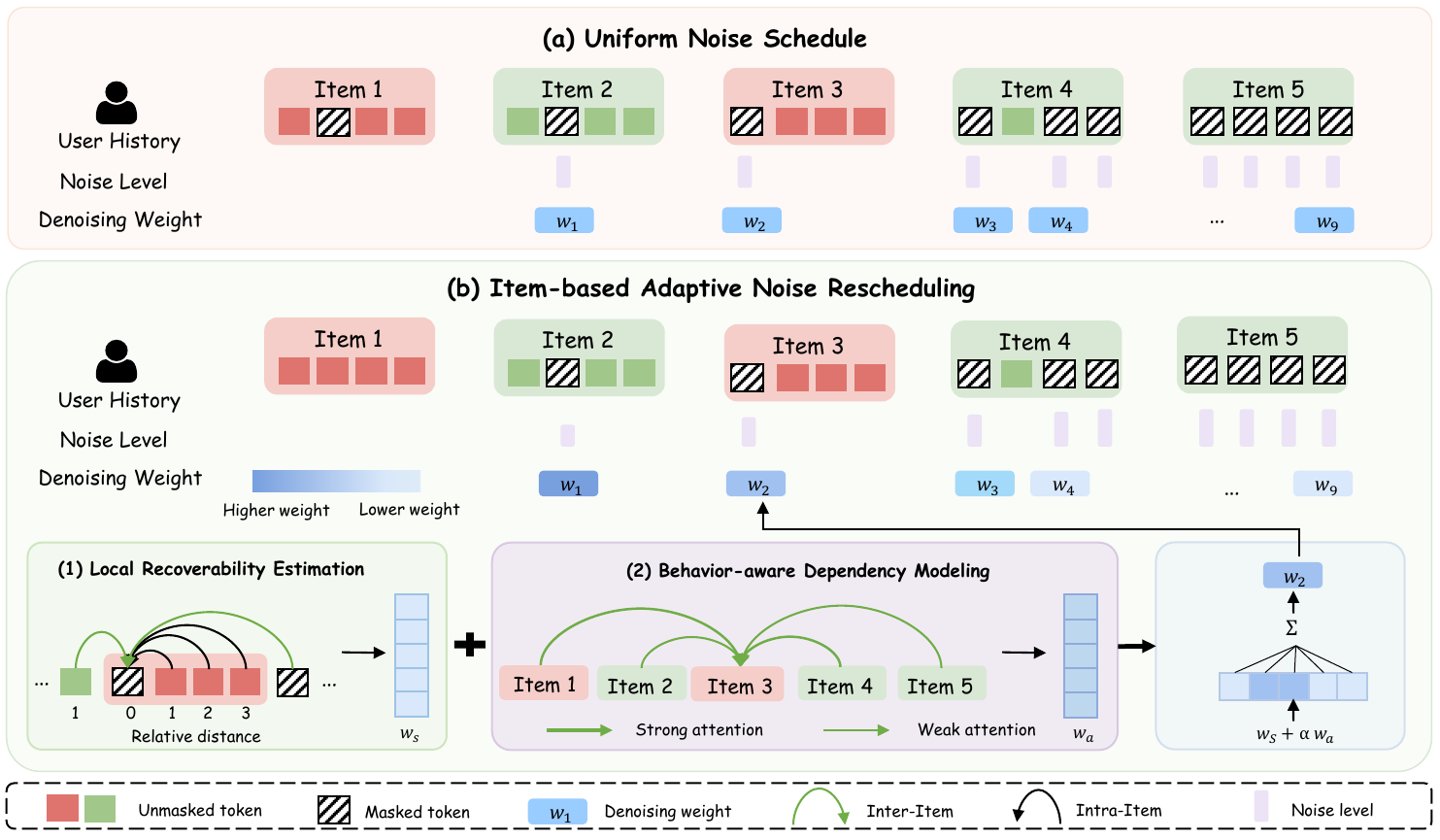}
    
    \caption{Illustration of the proposed item-based adaptive noise rescheduling strategy. (a) Existing diffusion-based recommendation models employ a uniform noise schedule, assigning identical denoising weights to masked tokens regardless of their contextual dependencies. (b) Our method dynamically adjusts the denoising weight for each token by jointly considering two complementary signals: (1) local recoverability estimation, and (2) behavior-aware dependency modeling. The final denoising weight is obtained by combining these two signals, enabling structure-aware diffusion training. }
    \label{fig:adpative_method}
\end{figure}

The recent surge in Large Language Models (LLMs) has revolutionized natural language processing~\cite{achiam2023gpt,liu2024deepseek,yang2025qwen3,zhao2023survey}, inspiring a paradigm shift in recommender systems towards generative recommendation~\cite{tang2025think,lin2025order,liu2025generative,liu2026cdrec,hu2025prefergrow}. Unlike traditional discriminative approaches that predict scores for a candidate set via point-wise ranking~\cite{chen2020bias,wang2023diffusion}, the generative paradigm reconceptualizes it as a sequence-to-sequence generation task, which is analogous to next-token prediction in autoregressive language modeling paradigms.

The paradigm of generative recommendation generally hinges on two fundamental pillars: item tokenization~\cite{hua2023index} and generation~\cite{zheng2024adaptinglcrec}. The first component, item tokenization, maps each item to a sequence of Semantic IDs (SIDs). This is typically achieved through methods such as hierarchical clustering or residual quantization (e.g., RQ-VAE), which capture semantic relationships within the ID structure. After the SIDs are generated, a generative model is applied to predict the target item tokens based on the user history SID sequence.

Discrete diffusion provides a natural generation paradigm for SID-based recommendation~\cite{liu2026diffgrm,shi2025lladarec}. Compared with autoregressive models that generate SID tokens in a fixed left-to-right order, discrete diffusion reconstructs masked tokens from partially corrupted sequences, enabling bidirectional context modeling and adaptive-order generation. This property is particularly suitable for recommendation, where the tokens in an SID jointly specify an item and may exhibit different semantic granularity and predictability. 

While the adaptation of discrete diffusion models to generative recommendation has yielded promising results, we argue that a fundamental bottleneck remains: the failure to explicitly encode the item-based collaborative filtering information within the generative framework.
Current DDM-based approaches primarily focus on modeling \textit{user-based} sequential patterns via iterative token denoising. However, they largely overlook the intrinsic \textit{item-based} structural relationships, the core signal in classic collaborative filtering.
This issue creates a disconnect between the generic generation paradigm and the specific requirements of recommendation, manifesting in two critical limitations where item-level structures are neglected in both item representation and the training objective.

The first limitation stems from the lack of collaborative signals in item tokenization. Existing discrete diffusion-based approaches predominantly rely on semantic embeddings (e.g., derived from textual descriptions) to construct SIDs. While effective at capturing content similarity, these representations are oblivious to the collaborative item relationships that are pivotal for SID-based recommendation~\cite{xiao2025unger,zheng2024adaptinglcrec}. For instance, two semantically distinct items may frequently co-occur in user histories, a pattern that standard SIDs fail to encode. Consequently, the generative model lacks a structured collaborative prior, forcing it to infer these item associations solely from sparse sequence contexts without support from the item representation.

Nevertheless, such a collaborative prior is insufficient, since different masked tokens may receive different amounts of contextual support under different corruption patterns. The second limitation stems from the uniform discrete diffusion training process, which cannot explicitly model item-based collaborative signals. As illustrated in Figure~\ref{fig:adpative_method}, existing DDM-based recommendation models generally adopt a uniform noise schedule, assigning the same noise level and denoising weight (e.g., $w_1=w_2=\cdots=w_9$) to all masked tokens regardless of their context. However, recommendation sequences naturally exhibit uneven information density. The recoverability of a masked token largely depends on the availability of related collaborative contexts, including both neighboring items in the sequence and informative tokens within the same semantic ID. Under a uniform schedule, structurally supported tokens and context-isolated tokens are treated equally during training, making it difficult for the model to effectively exploit item-level collaborative structures and behavioral dependencies.

To bridge the gap, we propose \ourname, a unified framework designed to systematically integrate item-based collaborative structures into generative recommendation.
To address the first challenge and provide a structural prior for the adaptive noise scheduler, we enhance Semantic ID construction by explicitly incorporating collaborative structural signals. Specifically, we construct an item-item co-occurrence matrix from historical interaction sequences and apply matrix factorization to derive structural item representations. These collaborative embeddings are then fused with pre-trained textual semantic features and discretized into hierarchical token sequences. By mapping items into this structurally-aware semantic token space, the resulting SIDs capture both semantic similarity and implicit transition patterns, thereby providing more informative conditions for the discrete diffusion denoising process.

To address the second challenge, we propose an item-based adaptive noise rescheduling mechanism for diffusion-based recommendation. As illustrated in Figure~\ref{fig:adpative_method}, instead of applying identical denoising weights to all masked positions, our method dynamically adjusts the denoising weight of each token according to two complementary signals: (1) \textit{Local Recoverability Estimation}, which estimates how easily a masked token can be recovered based on geometric dependencies from surrounding intra-item and inter-item contexts; and (2) \textit{Behavior-aware Dependency Modeling}, which leverages attention-based interaction modeling to adaptively capture collaborative dependencies in the user behavior sequence. By jointly integrating these two signals, our method assigns larger denoising weights to tokens supported by more informative structural and behavioral contexts, thereby providing fine-grained guidance for the diffusion denoising process. For clarity, the denoising weight shown in Figure~\ref{fig:adpative_method}(b) is computed for the masked token in Item 3. The figure demonstrates how surrounding unmasked contexts and behavior-aware item dependencies jointly influence the adaptive denoising weight assignment for the token. The denoising weight can be interpreted as an inverse effective noise level.  Tokens receiving larger denoising weights are considered easier to recover, corresponding to lower noise levels. In contrast, tokens with sparse contextual information are assigned smaller weights, indicating higher noise levels.
In summary, our main contributions are formulated as follows:
\begin{itemize}
   \item We propose an item-based adaptive noise rescheduling strategy in discrete diffusion that enables dynamic learning of item-to-item collaborative structures for generative recommendation.
   \item We introduce an explicit item co-occurrence–guided semantic ID generation mechanism, which serves as a lightweight collaborative prior to stabilize diffusion training.
    \item Extensive experiments across five diverse, real-world benchmark datasets demonstrate that \ourname~ consistently and significantly outperforms state-of-the-art autoregressive and diffusion-based generative recommendation models.
\end{itemize}

\section{Related Works}

\subsection{Generative Recommendation}
Unlike traditional discriminative approaches that formulate recommendations as a ranking task over a predefined candidate set~\cite{he2017neural, kang2018selfsteamsasrec,sun2019bert4rec,hidasi2015sessionGRU4Rec,yue2024linearlrurec,zhou2022filterfmlprec,li2023diffurec,yang2023generatedreamrec}, generative recommendation reframes the problem as a sequence-to-sequence generation task~\cite{si2024generative,zhai2024actions,geng2022recommendation,cui2022m6,liu2026cdrec}. In this paradigm, predicting the next item is isomorphic to next-token prediction in language modeling, unlocking the potential to leverage Large Language Models (LLMs) for recommendation.

Recent research has predominantly shifted towards generative recommendation with semantic IDs~\cite{wang2024eager,li2025dimerec,hu2025prefergrow} to bypass the inefficiency of generating lengthy natural language text. These works typically employ residual quantization (e.g., RQ-VAE~\cite{lee2022autoregressive}) or hierarchical clustering to convert atomic item IDs into highly compressed, discrete token sequences. TIGER~\cite{rajput2023recommendertiger} establishes this paradigm by generating tuple-based SIDs, enabling Transformers to predict items autoregressively. VQ-Rec~\cite{hou2023learningvqrec} parallelly enriches this approach by deriving discrete codes from item text to bridge content semantics. Building on these foundations, subsequent works such as LC-Rec~\cite{zheng2024adaptinglcrec} and LETTER~\cite{wang2024learnableletter} refine the tokenization mechanism through learnable codebooks and hierarchical regularization. Most recently, advanced frameworks like OneRec~\cite{deng2025onerec} and RPG~\cite{hou2025generatingrpg} unify multi-modal signals and optimize retrieval strategies for efficient generation.
Despite these advancements, the dominant autoregressive (AR) models rely heavily on a strict left-to-right generation order. This AR nature introduces two critical bottlenecks: (1) \text{Unidirectional Bias}, which restricts the model's ability to leverage bidirectional contextual semantics crucial for user intent understanding; and (2) \text{Error Propagation}, where a misprediction at an early hierarchical token (e.g., the coarse-grained cluster center) may propagate errors to subsequent token generations, leading to completely irrelevant item retrievals. 

To circumvent the limitations of autoregressive decoding, LLaDA-Rec~\cite{shi2025lladarec} and DiffGRM~\cite{liu2026diffgrm} introduce the discrete diffusion model to recommendation, enabling a non-autoregressive, bidirectional generation process. However, these DDM-based methods inherently rely on a uniform noise strategy, where tokens are masked stochastically without accounting for the density of surrounding contextual information. This denoising process fails to distinguish between easy-to-recover tokens and hard ones, severely limiting its ability to capture collaborative filtering signals through structure-aware guidance. Furthermore, relying solely on semantic-based tokenization lacks structured collaborative priors, further hindering the robust modeling of inter-item dependencies under noisy contexts.

\subsection{Discrete Diffusion Models}
Denoising Diffusion Probabilistic Models (DDPMs)~\cite{ho2020ddpm,yang2023diffusion} have achieved state-of-the-art performance in continuous state-space generation tasks like image synthesis. To address the discrete and categorical nature of language and recommendation data, Discrete Diffusion Models~\cite{shi2024simplified,zhu2025llada,yang2025mmada,ye2025dreamvl,xie2025dream} operate directly on categorical variables~\cite{ghazvininejad2019mask}. D3PM~\cite{austin2021d3pm} explicitly generalizes the diffusion framework to discrete state spaces using transition matrices, introducing absorbing states (e.g., \texttt{[MASK]}) to model the controlled corruption and progressive denoising processes. 

Building on these foundations, recent works have significantly scaled up this paradigm to the LLM era. LLaDA~\cite{nie2025llada} demonstrates that masked diffusion models trained from scratch can achieve performance highly competitive with strong autoregressive methods. Parallelly, frameworks like bd3lm~\cite{arriola2025block} and Dream~\cite{ye2025dream} further improve generation efficiency and quality through block-wise token updates and refined loss reweighting objectives. 
In the context of recommendation systems, the discrete diffusion model offers a compelling alternative to AR models~\cite{shi2025lladarec,liu2026diffgrm}. By training a bidirectional Transformer to reconstruct masked tokens from a partially corrupted sequence, it inherently learns to utilize global user history. However, SIDs in recommendation possess both intra-item and inter-item structures, making them fundamentally different from natural language text. 
Our work builds upon the discrete diffusion foundation but fundamentally diverges by introducing an item-based adaptive noise rescheduling mechanism. By integrating dynamic attention and structural priors, our approach is specifically tailored for collaborative SIDs, distinguishing it from general text diffusion models.

\section{Preliminaries and Background}
\subsection{Problem Formulation}
Consider a standard recommendation scenario involving a user set $\mathcal{U}$ and an item set $\mathcal{I}$. For each user $u \in \mathcal{U}$, we denote their interaction sequence as $\mathcal{H}_u = \{v_1, v_2, \dots, v_{|\mathcal{H}_u|}\}$, ordered chronologically, where $v_k \in \mathcal{I}$. The primary objective is to estimate the probability distribution of the subsequent item $v_{next}$ given the historical context $\mathcal{H}_u$.

Within the generative recommendation paradigm, items are not treated as atomic ID embeddings but are represented as sequences of discrete tokens. Specifically, each item $v$ is tokenized into a semantic ID, denoted as a vector of length $L$: $\mathbf{x}_v = [x_{v,1}, x_{v,2}, \dots, x_{v,L}]$, where each $x_{v,l}$ represents a discrete token derived from a quantization codebook. Consequently, a user's interaction history can be flattened into a unified token sequence $S_u$, formed by concatenating the SIDs of interacted items:
\begin{equation}
    S_u = [\mathbf{x}_{v_1} \oplus \mathbf{x}_{v_2} \oplus \cdots \oplus \mathbf{x}_{v_{|\mathcal{H}_u|}}],
\end{equation}
where $\oplus$ denotes sequence concatenation. The next-item prediction task is thus reframed as maximizing the conditional likelihood of the target item's token sequence $\mathbf{x}_{n}$ given the context $S_u$:
\begin{equation} 
    \Theta = \mathop{\arg\max}_{\theta} P_{\theta}(\mathbf{x}_{n} \mid S_u),
\end{equation}
where $\theta$ represents the learnable parameters of the model.

\subsection{Discrete Diffusion Modeling}
\label{sec:discrete_diffusion}
Diverging from the conventional Autoregressive approach that generates tokens from left to right, Discrete Diffusion operates as a non-autoregressive refinement process. It models generation as a reverse denoising trajectory, initiating from a completely corrupted state and progressively recovering the clean data.
Let $\mathbf{x}_n^T$ be the fully corrupted state of the next item SID consisting entirely of \texttt{[MASK]} tokens. Inference is discretized into $T$ steps. Starting from  $\mathbf{x}_n^T$, the Transformer encoder denoising network iteratively predicts the clean tokens.
At each step $t$ (from $T$ down to $1$), the denoising network takes the current state $\mathbf{x}_n^t$ of the next item and the unmasked user interaction history $S_H$ as input to predict the clean tokens for all masked positions.
In each iteration, the model predicts the probability distribution $P_{\theta}(\mathbf{x}_n^0 | \mathbf{x}_n^t, S_H)$. Let $n_t \in [1, L]$ denote the number of tokens to be kept at step $t$. We select the top-$n_t$ tokens with the highest prediction confidence and re-mask the remaining $L-n_t$ positions to form the input for the next step $\mathbf{x}_n^{t-1}$. Formally, the generation process can be viewed as maximizing the conditional likelihood over the refinement trajectory:
\begin{equation}
    P_{\theta}(\mathbf{x}_{n} \mid S_{H}) = \prod_{t=1}^{T} \prod_{l=1}^{L} 
    \begin{cases} 
    P_{\theta}(x_{n,l} \mid \mathbf{x}_n^t, S_{H}), & \text{if } x^t_{n,l} = \texttt{[MASK]}, \\
    1, & \text{otherwise}.
    \end{cases}
\end{equation}
Here, $x^t_{n,l}$ denotes the token at position $l$ in the state $\mathbf{x}_n^t$. 

\section{Methodology}
\label{sec:method}
\begin{figure*}
    \centering
    \includegraphics[width=1\linewidth]{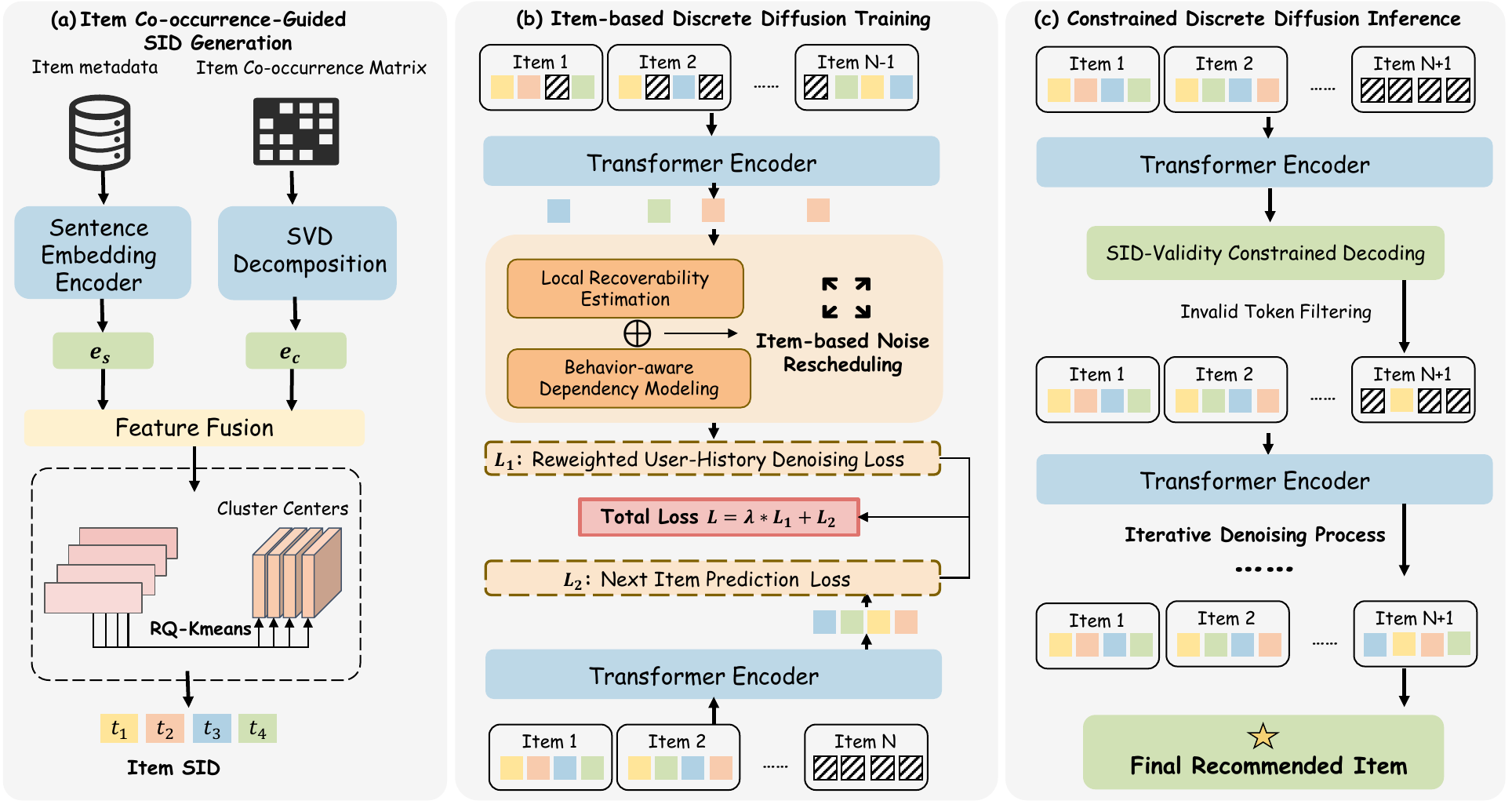}
    \caption{The architecture of \ourname. (a) SID generation via fusion of semantic and collaborative signals. (b) Discrete diffusion training enhanced by item-based adaptive noise rescheduling, enabling structure-aware denoising to capture collaborative filtering signals. (c) Constrained discrete diffusion inference with validity filtering.}
    \label{fig:model} 
\end{figure*}
In this section, we present our proposed framework. As illustrated in Figure~\ref{fig:model}, the framework consists of three components: (1) Item Co-occurrence-Guided Semantic ID Generation (Section~\ref{sec:sidcons}): we construct SIDs by fusing item textual information with collaborative signals; (2) Item-based Discrete Diffusion Training (Section~\ref{sec:ddmtrain}): we introduce an item-based adaptive noise rescheduling mechanism and employ a Transformer encoder to model the generation process; (3) Constrained Discrete Diffusion Inference (Section~\ref{sec:inference}): we employ iterative denoising with beam search, augmented by an SID constraint module to ensure the validity of the generated token sequence.

\subsection{Item Co-occurrence-Guided SID Generation}
\label{sec:sidcons}
\subsubsection{Item Representation Fusion}
Let $\mathcal{I}$ denote the set of items. For each item $i \in \mathcal{I}$, we construct a continuous representation by jointly encoding semantic and collaborative information. We first obtain a semantic embedding $\mathbf{e}_i^{\mathrm{sem}} \in \mathbb{R}^{d_s}$ using a pre-trained sentence embedding encoder. To fuse collaborative signals, we construct an item co-occurrence matrix $\mathbf{C} \in \mathbb{R}^{|\mathcal{I}| \times |\mathcal{I}|}$, where $\mathbf{C}_{ij}$ denotes the number of times items $i$ and $j$ co-occur in user interaction sequences. To extract low-dimensional collaborative features, we perform truncated singular value decomposition (SVD) on $\mathbf{C}$:
\begin{equation}
\mathbf{C} \approx \mathbf{U}_k \mathbf{\Sigma}_k \mathbf{V}_k^\top ,
\end{equation}
where $\mathbf{U}_k \in \mathbb{R}^{|\mathcal{I}| \times k}$ contains the top-$k$ left singular vectors.
The collaborative embedding of item $i$ is given by:
\begin{equation}
\mathbf{e}_i^{\mathrm{coll}} = \mathbf{U}_k(i,:) \mathbf{\Sigma}_k . 
\end{equation}

The final fused item representation is obtained by concatenation:
\begin{equation}
\mathbf{e}_i = [\, \mathbf{e}_i^{\mathrm{sem}} \, ; \, \mathbf{e}_i^{\mathrm{coll}} \,] \in \mathbb{R}^{d_s + d_c}.
\end{equation}
This fusion embedding preserves both semantic similarity and item collaborative proximity.

\subsubsection{RQ-KMeans-based SID Generation}
Given the fused item representations $\{\mathbf{e}_v\}_{v \in \mathcal{I}}$, we apply Residual Quantized K-Means (RQ-KMeans) to obtain SIDs. Each item is encoded as a sequence of $L$ discrete tokens.
RQ-KMeans performs multi-stage vector quantization in a residual manner. At stage $l=1$, K-Means is applied to cluster $\mathbf{e}_v$ into $|C_1|$ centroids, producing the first token:
\begin{equation}
    x_{v,1} = \mathop{\arg\min}_{\mathbf{c} \in C_1} \|\mathbf{e}_v - \mathbf{c}\|_2^2.
\end{equation}
The residual is then computed as: $\mathbf{r}_{v,1} = \mathbf{e}_v - \mathbf{c}_{x_{v,1}}.$
Subsequent stages recursively quantize the residual:
\begin{equation}
    x_{v,l} = \mathop{\arg\min}_{\mathbf{c} \in C_l} \|\mathbf{r}_{v,l-1} - \mathbf{c}\|_2^2, \quad \mathbf{r}_{v,l} = \mathbf{r}_{v,l-1} - \mathbf{c}_{x_{v,l}}.
\end{equation}
After $L$ stages, each item $v$ is represented as a SID:
\begin{equation}
    \text{SID}(v) = (x_{v,1}, x_{v,2}, \dots, x_{v,L}).
\end{equation}
By incorporating collaborative signals into the input space of RQ-KMeans, the resulting SIDs preserve both semantic and co-occurrence structures, yielding more informative corrupted contexts and an effective collaborative prior for discrete diffusion training.

\subsection{Item-based Discrete Diffusion Training}
\label{sec:ddmtrain}
Given a user interaction history consisting of $N-1$ items, each represented by $L$ SID tokens, the user history token sequence is denoted by:
\begin{equation}
    S_H = (x_{1,1}, \dots x_{1,L}, \dots,x_{N-1,1} \dots x_{N-1,L}), 
\end{equation}
The diffusion training process consists of two stages: forward masking and reverse reconstruction. In the forward process, tokens in the input sequence are independently replaced with a \texttt{[MASK]} token with probability $\tau \in (0, 1]$. In the reverse process, the model learns to reconstruct the original sequence from the noisy state by predicting masked tokens. We adopt two masking learning strategies for discrete diffusion training: user-history level masking and next-item level masking.

\subsubsection{User-History Level Masking}
User-history level masking aims to capture contextual dependencies among all tokens in the user history.
During training, for each batch, we sample a masking ratio $\tau \sim \mathcal{U}(0, 1]$. Each token in $S_H$ is independently masked with probability $\tau$, producing a partially corrupted sequence $S_H^\tau$. Let $\mathbb{I}[S_{H,i}^\tau = \texttt{[MASK]}]$ be an indicator function denoting whether the $i$-th token in $S_H$ is masked. The reconstruction loss is defined as:
\begin{small}  
\begin{equation}
    \mathcal{L}_{\text{UH}} = -\mathbb{E}_{\tau, S_H, S_H^\tau} \left[ \frac{1}{\tau} \sum_{i=1}^{(N-1) \times L} \mathbb{I}[S_{H,i}^\tau = \texttt{[MASK]}] \log p_{\theta}(S_{H,i} \mid S_H^\tau) \right].
\end{equation}
\end{small}

\subsubsection{Item-based Adaptive Noise Rescheduling}
To effectively capture item-based collaborative filtering signals during the generative process, we propose an item-based adaptive noise rescheduling strategy for user-history level masking. Unlike uniform denoising objectives, our mechanism quantifies the context of each token by jointly considering its structural hierarchy and dynamic user behavior.
Specifically, the contextual dependency between token $i$ and token $j$ is modeled as the combination of a local recoverability estimation and a behavior-aware item dependency score. 

First, we define a static structural dependency matrix $\mathbf{W}^{struct}$ based on symmetric geometric decay:
\begin{equation}
\label{eq:struct_weight}
W^{struct}_{ij}=\phi(|I(i)-I(j)|)+\mathbb{I}[I(i)=I(j)]\cdot\psi(|T(i)-T(j)|),
\end{equation}
where $I(i)$ denotes the item index associated with token $i$, and $T(i)$ denotes its intra-item token position. The functions $\phi(d)$ and $\psi(d)$ are geometric decay kernels defined as
\[p(1-p)^{d-1}, \quad d\ge 1,\]
which model inter-item and intra-item contextual recoverability, respectively. Here, $d$ is the distance, and $p\in(0,1]$ controls the decay sharpness: a larger $p$ emphasizes nearby contextual information, while a smaller $p$ captures broader contextual dependencies. We set $\phi(0)=0$ and $\psi(0)=0$ to avoid self-dependency.

However, purely relying on static structural context cannot adaptively capture user-specific behavioral dependencies. To address this limitation, we further introduce behavior-aware item dependency modeling based on item-level attention. 
Specifically, let $\mathbf{e}_m \in \mathbb{R}^d$ denote the representation of the $m$-th item in the user sequence, obtained by mean pooling the hidden states of its corresponding tokens from the last Transformer layer. To avoid degenerate optimization, we detach $\mathbf{e}_m$ from the computational graph when computing the reweighting coefficients. This design prevents the model from trivially reducing the objective by manipulating the adaptive weights. We compute the item-level attention score between item $m$ and item $n$ as:
\begin{equation}
\label{eq:item_attention}
A^{item}_{mn}
=\frac{
\exp(
\mathbf{e}_m^\top
\mathbf{W}_Q^\top
\mathbf{W}_K
\mathbf{e}_n
/\sqrt{d}
)
}{
\sum_{k}
\exp(
\mathbf{e}_m^\top
\mathbf{W}_Q^\top
\mathbf{W}_K
\mathbf{e}_k
/\sqrt{d}
)
},
\end{equation}
where $\mathbf{W}_Q$ and $\mathbf{W}_K$ are learnable projection matrices.
The item-level dependency score is then projected back to token pairs according to their corresponding item indices:
\begin{equation}
\label{eq:token_attention}
A_{ij}=A^{item}_{I(i),I(j)}.
\end{equation}

Finally, the adaptive contextual dependency weight is obtained by combining the structural context with the behavior-aware dependency score:
\begin{equation}
\label{eq:final_weight}
W_{ij}
=W^{struct}_{ij}+\alpha A_{ij},
\end{equation}
where $\alpha \geq 0$ is a learnable scaling parameter. This formulation encourages stronger denoising guidance when two tokens are either structurally proximate or semantically correlated through user behavioral patterns.
To estimate the denoising importance of each masked token, we aggregate contextual contributions from all unmasked tokens:
\begin{small}
\begin{equation}
\mathbf{W}_{i}
=\sum_{j=1}^{(N-1)\times L}\mathbb{I}[S_{H,j}^\tau \neq \texttt{[MASK]}] \cdot W_{ij}.
\end{equation}
\end{small}

The reweighted denoising objective is formulated as:
\begin{small}
\begin{equation}
\mathcal{L}_{1}
=
-\mathbb{E}_{\tau,S_H,S_H^\tau}
\left[
\frac{1}{\tau}
\sum_{i=1}^{(N-1)\times L}
\mathbf{W}_{i}
\cdot
\mathbb{I}[S_{H,i}^\tau=\texttt{[MASK]}]
\log p_\theta(S_{H,i}\mid S_H^\tau)
\right].
\label{eq:uh_car_loss}
\end{equation}
\end{small}

This adaptive noise rescheduling strategy redistributes denoising importance according to both local contextual recoverability and behavior-aware item dependencies. Tokens surrounded by richer unmasked contexts or strongly correlated items receive larger denoising weights, corresponding to lower effective noise levels during training. In contrast, tokens with sparse contextual support are assigned smaller weights. By explicitly aligning denoising supervision with item-level collaborative structures, the proposed strategy provides fine-grained and structure-aware guidance for discrete diffusion training.

\subsubsection{Next-Item Level Masking}
Next-item level masking focuses on predicting the target item, conditioned on the fully observed user history.
We define $\mathbf{x}_{n} = (x_{n,1}, \dots, x_{n,L})$ as the SIDs of the next item.
Given mask ratio $\tau$, each token in $\mathbf{x}_{n}$ is independently masked with probability $\tau$, yielding $\mathbf{x}_{n}^\tau$.
The partially masked target item tokens are concatenated with the unmasked user history $S_H$ and fed into the model.
The next-item prediction loss is defined as:
\begin{small}
\begin{equation}
\mathcal{L}_{{2}}
=
-\mathbb{E}_{\tau,\, S_H,\, \mathbf{x}_{n},\, \mathbf{x}_{n}^\tau}
\left[
\frac{1}{\tau}
\sum_{i=1}^{L}
\mathbb{I}[x_{n,i}^\tau = \texttt{[MASK]}]
\log p_\theta(x_{n,i} \mid S_H, \mathbf{x}_{n}^\tau)
\right].
\label{eq:ni_loss}
\end{equation}
\end{small}

\subsubsection{Overall Training Objective}
The final training objective combines the two masking strategies:
\begin{equation}
\label{eq:union_loss}
\mathcal{L}
=\lambda\mathcal{L}_{1}+\mathcal{L}_{2},
\end{equation}
where $\lambda$ balances user history modeling and target item generation.

\subsection{Constrained Discrete Diffusion Inference}
\label{sec:inference}
At inference time, our model generates the next item conditioned on the observed user history by progressively denoising a fully masked SID.

\subsubsection{Initialization}
Given a user history sequence $S_H$, we initialize the target item SID as a fully masked sequence $\textbf{x}_n^{T} = (x_{n,1}^{T}, \dots, x_{n,L}^{T})$, $\quad x_{n,m}^{T} = \texttt{[MASK]},$ and concatenate it with $S_H$ as the input to the denoising network, implemented as a Transformer encoder.
\subsubsection{Iterative Denoising with Beam Search.}
Inference proceeds over $T$ diffusion steps. At each step $t$, the model predicts token distributions for the masked positions:
\begin{equation}
p_\theta(x_{n,m} \mid S_H, \mathbf{x}_n^{t}).
\end{equation}
Following LLaDA-Rec~\cite{shi2025lladarec}, we adopt a beam search strategy to maintain multiple candidate partially denoised sequences.
At each step, candidate sequences in the beam are expanded by unmasking a subset of positions and selecting high-probability tokens.
The beam retains the top-$B$ candidates according to their cumulative log-likelihoods.
\subsubsection{SID-Validity Constrained Decoding.}
In discrete diffusion, SIDs may be revealed in an arbitrary order. Enforcing validity solely at the level of individual positions is insufficient, as independently valid token assignments may form an invalid combination. To guarantee that the final generated SID corresponds to a real item, we enforce an SID-validity constraint during decoding.

During inference, we maintain a candidate item set $\mathcal{C}^{t} \subseteq \mathcal{I}$ at diffusion step $t$, representing all items that remain consistent with the currently revealed tokens.
Formally, let $\mathcal{O}^{t} = \{(m, {x}_{n,m}) \mid {x}_{n,m} \neq \texttt{[MASK]}\}$ denote the set of observed (position, token) pairs at step $t$.
The candidate set is defined as:
\begin{equation}
\mathcal{C}^{t} =
\left\{
i \in \mathcal{I}
\;\middle|\;
\forall (m, {x}_{n,m}) \in \mathcal{O}^{t},\;
\mathrm{SID}(i)[m] = {x}_{n,m}
\right\}.
\end{equation}
When unmasking a token at position $m$, the valid candidate token set is restricted to:
\begin{equation}
\mathcal{X}_m^{t} =
\left\{
x \;\middle|\;
\exists\, i \in \mathcal{C}^{t}
\text{ such that }
\mathrm{SID}(i)[m] = x
\right\}.
\end{equation}
The predictive distribution is renormalized over $\mathcal{X}_m^{t}$ before token selection. After assigning a value to position $m$, the candidate set is updated accordingly. 

\subsubsection{Final Prediction.}
After all $L$ tokens are unmasked, each beam candidate corresponds to a valid SID. The candidate with the highest overall likelihood is selected, and its corresponding item is returned as the final recommendation.

To consolidate the aforementioned components and provide a clear implementation overview, we summarize the overall workflow of \ourname. The complete training pipeline is outlined in Algorithm \ref{alg:training}. Correspondingly, the constrained iterative decoding process used during the inference phase is detailed in Algorithm \ref{alg:inference}.

\begin{algorithm}[htbp]
\caption{Training Process of \ourname}
\label{alg:training}
\begin{algorithmic}[1]
\REQUIRE User interaction sequences $\mathcal{H}$, Items $\mathcal{I}$, Diffusion steps $T$, Masking ratio distribution $\mathcal{U}(0,1]$, Hyperparameter $\lambda, p$.
\ENSURE Trained network parameters $\theta$.
\STATE \textbf{Offline Preparation:}
\STATE Compute Item-Item co-occurrence matrix $\mathbf{C}$.
\STATE Apply SVD to obtain collaborative embeddings and fuse with semantic embeddings.
\STATE Generate semantic IDs using RQ-KMeans.
\STATE \textbf{Online Training:}
\WHILE{not converged}
    \STATE Sample a batch of user interaction histories $\mathcal{S}_H$ and target items $\mathbf{x}_n$.
    \STATE Sample masking ratios $\tau_1, \tau_2 \sim \mathcal{U}(0,1]$.
    \STATE \textit{// User-History Level Masking}
    \STATE Mask tokens in $\mathcal{S}_H$ with probability $\tau_1$. 
    \STATE Extract item-level hidden states $\mathbf{e}$ from the last Transformer layer.
    \STATE Compute dynamic  attention scores $A_{ij}$ using $\mathbf{e}$.
    \STATE Compute structural recoverability weights $W^{struct}_{ij}$ based on geometric decay $p$ (Eq.~\ref{eq:struct_weight}).
    \STATE Calculate context-adaptive weights $W_{ij} = W^{struct}_{ij} + \alpha A_{ij}$.
    \STATE Compute reweighted history loss $\mathcal{L}_1$ (Eq.~\ref{eq:uh_car_loss}).
    \STATE \textit{// Next-Item Level Masking}
    \STATE Mask tokens in target item $\mathbf{x}_n$ with probability $\tau_2$.
    \STATE Compute target item prediction loss $\mathcal{L}_2$ given unmasked $\mathcal{S}_H$ (Eq.~\ref{eq:ni_loss}).
    \STATE \textit{// Parameter Update}
    \STATE Compute total loss $\mathcal{L} = \lambda \mathcal{L}_1 + \mathcal{L}_2$.
    \STATE Update network parameters $\theta$ via gradient descent.
\ENDWHILE
\end{algorithmic}
\end{algorithm}

\begin{algorithm}[htbp]
\caption{Inference Process with Constrained Decoding}
\label{alg:inference}
\begin{algorithmic}[1]
\REQUIRE User history SIDs $\mathcal{S}_H$, Trained model $p_\theta$, Beam size $B$, Diffusion steps $T$.
\ENSURE Recommended target item $i^*$.
\STATE Initialize candidate target item $\mathbf{x}_n^T$ as fully [MASK] tokens.
\STATE Initialize beam with sequence $\mathbf{x}_n^T$ and cumulative log-likelihood $0$.
\STATE Initialize valid candidate item set $\mathcal{C}^T$.
\FOR{step $t = T$ \TO $1$}
    \STATE \textbf{Predict:} Pass $[\mathcal{S}_H ; \mathbf{x}_n^t]$ into $p_\theta$ to get token distributions .
    \FOR{each sequence in beam}
        \STATE Identify the position $m$ to unmask.
        \STATE \textbf{Constrain:} Filter invalid tokens by updating valid token set $\mathcal{X}_m^t$ based on $\mathcal{C}^t$.
        \STATE Expand sequence with top probability valid tokens.
    \ENDFOR
    \STATE \textbf{Prune:} Retain top-$B$ sequences based on cumulative log-likelihood.
    \STATE \textbf{Update Valid Set:} $\mathcal{C}^{t-1} \leftarrow$ items consistent with newly unmasked tokens.
\ENDFOR
\STATE \textbf{Return} item $i^*$ corresponding to the top-1 generated valid SID sequence.
\end{algorithmic}
\end{algorithm}
\section{Experiments}

\subsection{Experimental Setup}
\subsubsection{Datasets}
\begin{table}[t] 
  \centering
  \caption{Statistics of the used datasets. ``Avg.Len'' indicates the average number of interactions per user.}
  \label{tab:datasets}
  \small %
  \setlength{\tabcolsep}{4pt} %
  \begin{tabular}{l ccc cc} 
    \toprule
    \textbf{Dataset} & \textbf{\#Users} & \textbf{\#Items} & \textbf{\#Interactions} & \textbf{Sparsity} & \textbf{Avg.Len} \\
    \midrule
    Scientific & 50,985 & 25,848 & 412,947 & 99.969\% & 8.10 \\
    Instrument & 57,439 & 24,587 & 511,836 & 99.964\% & 8.91 \\
    Game       & 94,762 & 25,612 & 814,586 & 99.966\% & 8.60 \\
    Steam       & 39,795 & 9,265 & 2,949,605 & 99.200\% & 74.12 \\
    MovieLens       & 6,040 & 3,883 & 1,001,456 & 95.730\% & 165.57 \\
    \bottomrule
  \end{tabular}
\end{table}
We evaluate our method on five real-world datasets across different domains. From the Amazon collection~\cite{hou2024bridgingdata}, we select three subsets: Scientific, Musical Instruments, and Video Games. In addition to these product-based datasets, we include MovieLens~\cite{harper2015movielens}, which contains user ratings alongside movie genres and titles, and Steam~\cite{pathak2017generating}, a dataset comprising user-item interactions and textual reviews from the Steam gaming platform.
These datasets vary in scale and domain, providing a comprehensive benchmark for recommendations. We follow the standard data split as in previous generative recommendation works~\cite{rajput2023recommendertiger,hou2023learningvqrec}, where each user’s historical reviews are treated as interaction records and arranged in chronological order, with the earliest review appearing first. The detailed statistics of these datasets are summarized in Table~\ref{tab:datasets}. 
We evaluate our model using the widely adopted leave-one-out protocol~\cite{shi2025lladarec}. Formally, for a sequence of $n$ interactions, we reserve the $n$-th item for final performance evaluation and the $(n-1)$-th item for validation. All prior interactions from $1$ to $n-2$ are used to optimize the model. 

\subsubsection{Evaluation Metrics}
To quantify the recommendation quality, we employ two standard ranking-based metrics consistently used in prior literature ~\cite{shi2025lladarec}: Recall$@k$ and Normalized Discounted Cumulative Gain (NDCG$@k$). In our experiments, we report the performance across $k \in \{1, 5, 10\}$. Note that NDCG$@1$ yields the same numerical value as Recall$@1$.

\subsubsection{Baselines}
To comprehensively evaluate our proposed model, we compare it against 16 representative baselines. Based on their item representation strategies, we categorize these models into two main groups: traditional Item ID-based methods and recent Semantic ID-based generative methods.

\vspace{1mm}
\noindent\textbf{Item ID-based Baselines.} 
This group treats items as atomic IDs. The selected baselines encompass both classical sequential models and generative approaches:
\begin{itemize}
    \item \textbf{GRU4Rec}~\cite{hidasi2015sessionGRU4Rec} models chronological user action sequences using Gated Recurrent Units.
    \item \textbf{SASRec}~\cite{kang2018selfsteamsasrec} captures sequential dynamics through a unidirectional, left-to-right Transformer architecture with self-attention.
    \item \textbf{BERT4Rec}~\cite{sun2019bert4rec} utilizes a bidirectional Transformer trained via a Cloze objective to predict masked item IDs.
    \item \textbf{FMLP-Rec}~\cite{zhou2022filterfmlprec} achieves efficient sequence encoding by replacing standard self-attention mechanisms with learnable filter-enhanced MLPs.
    \item \textbf{LRURec}~\cite{yue2024linearlrurec} employs linear recurrent units to maintain linear-time training and inference complexity while overcoming the non-linearity limitations of traditional linear transitions.
    \item \textbf{DreamRec}~\cite{yang2023generatedreamrec} and \textbf{DiffuRec}~\cite{li2023diffurec} formulate recommendation as a continuous denoising generation process by representing item IDs as continuous embeddings.
    \item \textbf{PreferGrow}~\cite{hu2025prefergrow} is a discrete diffusion-based model that directly captures ranking dynamics by fading and growing user preference ratios over the discrete item corpus.
\end{itemize}

\vspace{1mm}
\noindent\textbf{Semantic ID-based Baselines.} 
This group represents items as Semantic IDs to enable generative recommendation. The evaluated models include:
\begin{itemize}
    \item \textbf{VQ-Rec}~\cite{hou2023learningvqrec} facilitates sequence-to-sequence recommendation by deriving discrete item codes from textual representations via Vector Quantization.
    \item \textbf{TIGER}~\cite{rajput2023recommendertiger} constructs tuple-based semantic IDs using RQ-VAE and predicts the next item autoregressively.
    \item \textbf{TIGER-SAS}~\cite{rajput2023recommendertiger} integrates the semantic IDs generated by TIGER into the standard SASRec architecture.
    \item \textbf{LETTER}~\cite{wang2024learnableletter} and \textbf{LC-Rec}~\cite{zheng2024adaptinglcrec} enhance generative recommendation through learnable tokenizers, allowing for the joint optimization of the codebook and the downstream recommendation objective.
    \item \textbf{RPG}~\cite{hou2025generatingrpg} introduces a lightweight recommendation model to parallelly generate long, unordered semantic IDs.
    \item \textbf{DiffGRM}~\cite{liu2026diffgrm} is a diffusion-based model that replaces the traditional autoregressive decoder with a masked discrete diffusion framework.
    \item \textbf{LLaDA-Rec}~\cite{shi2025lladarec} serves as a state-of-the-art discrete diffusion model, generating SIDs non-autoregressively through a mask-and-predict denoising framework.
\end{itemize}

\subsubsection{Implementation Details}

\textbf{SID Generation.} Each item is represented by concatenating textual and collaborative embeddings. Specifically, we leverage a pre-trained \text{Sentence-T5} model~\cite{ni2022sentence} to extract semantic features from item metadata (e.g., titles and descriptions). Concurrently, to capture structural interaction patterns, we build an item co-occurrence matrix and apply SVD to derive a 64-dimensional collaborative embedding for each item. The combined embeddings are then discretized into SIDs using RQ-KMeans with $L=4$ levels and a codebook size of $K=256$ per level.

\textbf{Discrete Diffusion Model.} The discrete diffusion model is configured as a bidirectional Transformer encoder (8 attention heads, 256-dimensional hidden states). We deploy a 4-layer architecture for the \textit{Scientific}, \textit{Instrument},  \textit{Steam}, and \textit{MovieLens} while extending it to 6 layers for the \textit{Game} dataset to accommodate its larger scale. The scaling factor $\lambda$ in Eq.~\ref{eq:union_loss} is determined through a grid search over $\{1, 2, 3, 4, 5\}$. $\lambda$ is determined to be 3, 5, 2, 3, and 3 for the Scientific, Instrument, Game, MovieLens, and Steam datasets, respectively. The symmetric-geometric noise rescheduling sharpness $p$ is determined through a grid search over $\{0.1,0.5,0.9\}$. We train the Transformer encoder for 150 epochs with an early stopping mechanism.  The AdamW~\cite{loshchilov2017decoupled} optimizer is used with a batch size of 1024, where the learning rate and weight decay are fine-tuned within $\{0.005, 0.003, 0.001\}$ and $\{0.05, 0.005, 0.001\}$, respectively.

Our framework is implemented in PyTorch and executed on a workstation featuring the \text{Intel Xeon Gold 6226R CPU} and two \text{NVIDIA RTX 4090} GPUs.

\subsection{Main Results}
\subsubsection{Overall Performance}
\begin{table}[t]
\centering
\caption{Performance comparison on three Amazon datasets. R@$K$ and N@$K$ denote Recall@$K$ and NDCG@$K$, respectively. The best results are highlighted in \textbf{bold}, and the second-best results are \underline{underlined}. The improvements of our model over the best baseline are statistically significant (paired t-test, $p < 0.05$).}
\label{tab:main}
\setlength{\tabcolsep}{2.5pt}
\resizebox{\textwidth}{!}{%
\begin{tabular}{@{}ll ccccc ccccc ccccc@{}}
\toprule
\multirow{2}{*}{Category} & \multirow{2}{*}{Model} & \multicolumn{5}{c}{Scientific} & \multicolumn{5}{c}{Instrument} & \multicolumn{5}{c}{Game} \\
\cmidrule(lr){3-7} \cmidrule(lr){8-12} \cmidrule(lr){13-17}
& & R@1 & R@5 & R@10 & N@5 & N@10 & R@1 & R@5 & R@10 & N@5 & N@10 & R@1 & R@5 & R@10 & N@5 & N@10 \\
\midrule
\multirow{8}{*}{\makecell[l]{Item ID\\based}} 
& GRU4Rec~\citeyearpar{hidasi2015sessionGRU4Rec} & 0.0071 & 0.0184 & 0.0272 & 0.0128 & 0.0156 & 0.0094 & 0.0297 & 0.0453 & 0.0196 & 0.0246 & 0.0149 & 0.0461 & 0.0712 & 0.0307 & 0.0387 \\
& SASRec~\citeyearpar{kang2018selfsteamsasrec} & 0.0063 & 0.0240 & 0.0379 & 0.0152 & 0.0197 & 0.0089 & 0.0331 & 0.0525 & 0.0211 & 0.0273 & 0.0128 & 0.0516 & 0.0823 & 0.0323 & 0.0421 \\
& BERT4Rec~\citeyearpar{sun2019bert4rec} & 0.0045 & 0.0157 & 0.0264 & 0.0100 & 0.0134 & 0.0065 & 0.0255 & 0.0412 & 0.0160 & 0.0211 & 0.0082 & 0.0315 & 0.0530 & 0.0199 & 0.0267 \\
& FMLP-Rec~\citeyearpar{zhou2022filterfmlprec} & 0.0046 & 0.0181 & 0.0300 & 0.0113 & 0.0151 & 0.0086 & 0.0299 & 0.0496 & 0.0193 & 0.0257 & 0.0099 & 0.0395 & 0.0649 & 0.0246 & 0.0328 \\
& LRURec~\citeyearpar{yue2024linearlrurec} & 0.0049 & 0.0169 & 0.0267 & 0.0110 & 0.0141 & 0.0071 & 0.0272 & 0.0431 & 0.0172 & 0.0223 & 0.0134 & 0.0480 & 0.0753 & 0.0308 & 0.0396 \\
& DreamRec~\citeyearpar{yang2023generatedreamrec} & 0.0052 & 0.0184 & 0.0299 & 0.0118 & 0.0155 & 0.0069 & 0.0245 & 0.0423 & 0.0157 & 0.0214 & 0.0125 & 0.0381 & 0.0611 & 0.0253 & 0.0326 \\
& DiffuRec~\citeyearpar{li2023diffurec} & 0.0050 & 0.0190 & 0.0310 & 0.0119 & 0.0158 & 0.0077 & 0.0283 & 0.0465 & 0.0179 & 0.0237 & 0.0111 & 0.0425 & 0.0709 & 0.0268 & 0.0359 \\
& PreferGrow~\citeyearpar{hu2025prefergrow} & 0.0065 & 0.0172 & 0.0283 & 0.0101 & 0.0125 & 0.0094 & 0.0265 & 0.0397 & 0.0181 & 0.0224 & 0.0180 & 0.0521 & 0.0784 & 0.0351 & 0.0436 \\
\midrule
\multirow{9}{*}{\makecell[l]{Semantic ID\\based}}
& VQ-Rec~\citeyearpar{hou2023learningvqrec} & 0.0076 & 0.0248 & 0.0385 & 0.0162 & 0.0206 & 0.0099 & 0.0345 & 0.0532 & 0.0222 & 0.0282 & 0.0150 & 0.0497 & 0.0769 & 0.0325 & 0.0412 \\
& TIGER~\citeyearpar{rajput2023recommendertiger} & 0.0084 & 0.0282 & 0.0446 & 0.0183 & 0.0236 & 0.0105 & 0.0359 & 0.0566 & 0.0233 & 0.0300 & 0.0166 & 0.0529 & 0.0823 & 0.0348 & 0.0442 \\
& TIGER-SAS~\citeyearpar{rajput2023recommendertiger} & 0.0067 & 0.0221 & 0.0356 & 0.0144 & 0.0187 & 0.0102 & 0.0342 & 0.0521 & 0.0223 & 0.0280 & 0.0170 & 0.0548 & 0.0847 & 0.0360 & 0.0457 \\
& LETTER~\citeyearpar{wang2024learnableletter} & 0.0082 & 0.0273 & 0.0423 & 0.0179 & 0.0227 & 0.0114 & 0.0362 & 0.0562 & 0.0239 & 0.0303 & 0.0169 & 0.0552 & 0.0863 & 0.0362 & 0.0462 \\
& LC-Rec~\citeyearpar{zheng2024adaptinglcrec} & 0.0091 & 0.0280 & 0.0434 & 0.0186 & 0.0235 & 0.0119 & 0.0379 & 0.0587 & 0.0251 & 0.0318 & 0.0165 & 0.0567 & 0.0891 & 0.0366 & 0.0471 \\
& RPG~\citeyearpar{hou2025generatingrpg} & 0.0087 & 0.0257 & 0.0395 & 0.0174 & 0.0218 & 0.0118 & 0.0362 & 0.0545 & 0.0241 & 0.0300 & \underline{0.0209} & 0.0579 & 0.0853 & 0.0397 & 0.0485 \\
& DiffGRM~\citeyearpar{liu2026diffgrm} & 0.0092 & 0.0279 & 0.0441 & 0.0189 & 0.0232 & 0.0125 & 0.0382 & 0.0593 & 0.0255 & 0.0316 & 0.0205 & 0.0602 & 0.0925 & 0.0401 & 0.0483 \\
& LLaDA-Rec~\citeyearpar{shi2025lladarec} & \underline{0.0098} & \underline{0.0310} & \underline{0.0474} & \underline{0.0203} & \underline{0.0256} & \underline{0.0128} & \underline{0.0406} & \underline{0.0623} & \underline{0.0268} & \underline{0.0337} & 0.0207 & \underline{0.0623} & \underline{0.0942} & \underline{0.0415} & \underline{0.0517} \\
& \textbf{\ourname} & \textbf{0.0122} & \textbf{0.0327} & \textbf{0.0494} & \textbf{0.0226} & \textbf{0.0279} & \textbf{0.0135} & \textbf{0.0410} & \textbf{0.0629} & \textbf{0.0273} & \textbf{0.0343} & \textbf{0.0214} & \textbf{0.0653} & \textbf{0.0987} & \textbf{0.0434} & \textbf{0.0542} \\
\bottomrule
\end{tabular}}
\end{table}
The experimental results on three Amazon datasets are summarized in Table~\ref{tab:main}. Additionally, to further validate the generalization of our method across different domains, we conduct experiments on the MovieLens and Steam datasets, as presented in Table \ref{tab:main_2}. For these two datasets, we benchmark our method against a selective subset of the most competitive generative state-of-the-art models (i.e., RPG and LLaDA-Rec) alongside recently proposed strong baselines (PreferGrow and DiffGRM). By comparing our proposed model with various baselines, we can draw the following conclusions:

\begin{itemize}
    \item \textbf{Superiority of \ourname:} Our model consistently achieves the best performance across all datasets and evaluation metrics. Notably, \ourname~significantly outperforms the strongest generative baseline based on the discrete diffusion model, LLaDA-Rec, across most metrics. For example, on the \textit{Scientific} dataset, \ourname~delivers a substantial improvement in Recall@1. This performance gain validates the effectiveness of our noise rescheduling strategy, which enables the model to capture item-based collaborative filtering signals more precisely. 
    \item \textbf{Semantic ID vs. Item ID:} SID-based generative methods (e.g., VQ-Rec, TIGER, and RPG) consistently outperform traditional item ID-based models such as SASRec and BERT4Rec across multiple benchmarks. This suggests that Semantic IDs provide a more informative representation space than independent atomic item IDs. Furthermore, incorporating collaborative structural signals via our item co-occurrence modeling further improves the ability of SIDs to capture item relationships and sequential transition patterns.
    \item \textbf{Effectiveness of Discrete Diffusion:} Among SID-based approaches, DDM-based models (LLaDA-Rec, DiffGRM and \ourname) demonstrate superior ranking capabilities compared to autoregressive-based generators. This suggests that the bidirectional attention is more robust for recommendation tasks. Our model further improves upon this by adaptively rescheduling the noise level based on the geometric properties of the context, improving item retrieval accuracy.
\end{itemize}

\begin{table}[t]
\centering
\caption{Performance comparison on MovieLens and Steam. The best results are highlighted in \textbf{bold}, and the second-best results are \underline{underlined}. The improvements of our model over the best baseline are statistically significant (paired t-test, $p < 0.05$).}
\setlength{\tabcolsep}{5pt}
\resizebox{0.6\textwidth}{!}{%
\begin{tabular}{l|l|cccc}
\hline
Dataset & Model 
& NDCG@5 & NDCG@10 
& Recall@5 & Recall@10 \\
\hline

\multirow{6}{*}{MovieLens}
& PreferGrow~\citeyearpar{hu2025prefergrow}
& 0.0910 & 0.1177 
& 0.1409 & 0.2237 \\

& RPG~\citeyearpar{hou2025generatingrpg}
& 0.1197 & 0.1440 
& 0.1750 & 0.2503 \\

& DiffGRM~\citeyearpar{liu2026diffgrm}
& 0.1038 & 0.1260 
& 0.1541 & 0.2235 \\

& LLaDA-Rec~\citeyearpar{shi2025lladarec}
&\underline{ 0.1359 }& \underline{0.1628} 
& \underline{0.1980} & \underline{0.2820} \\

& \textbf{ANR-DiffRec}
& \textbf{0.1468} & \textbf{0.1738} 
& \textbf{0.2141} & \textbf{0.2977} \\

\hline

\multirow{6}{*}{Steam}

& PreferGrow~\citeyearpar{hu2025prefergrow}
& \underline{0.0406} & \underline{0.0508} 
& \underline{0.0615} & \underline{0.0935} \\

& RPG~\citeyearpar{hou2025generatingrpg} 
& 0.0388 & 0.0490 
& 0.0579 & 0.0898 \\

& DiffGRM~\citeyearpar{liu2026diffgrm}
&0.0378 & 0.0479
& 0.0565 & 0.0870 \\

& LLaDA-Rec~\citeyearpar{shi2025lladarec}
& 0.0391 & 0.0502 
& 0.0607 & 0.0914 \\

& \textbf{ANR-DiffRec} 
& \textbf{0.0411} & \textbf{0.0526} 
& \textbf{0.0627} & \textbf{0.0991} \\
\hline
\end{tabular}
}
\label{tab:main_2}
\end{table}

\subsubsection{Performance on Sequence Length}
To investigate the robustness of our model against varying amounts of user historical information, we categorize the users into four groups based on their interaction sequence lengths: short ($<5$), medium ($5 \sim 10$), long ($10 \sim 20$), and extremely long ($>20$). We evaluate NDCG@5 and Recall@5 of \ourname~ alongside two strong baselines: the autoregressive state-of-the-art RPG, and the diffusion-based LLaDA-Rec.

\begin{figure}[htbp]
    \centering
    \includegraphics[width=0.75\linewidth]{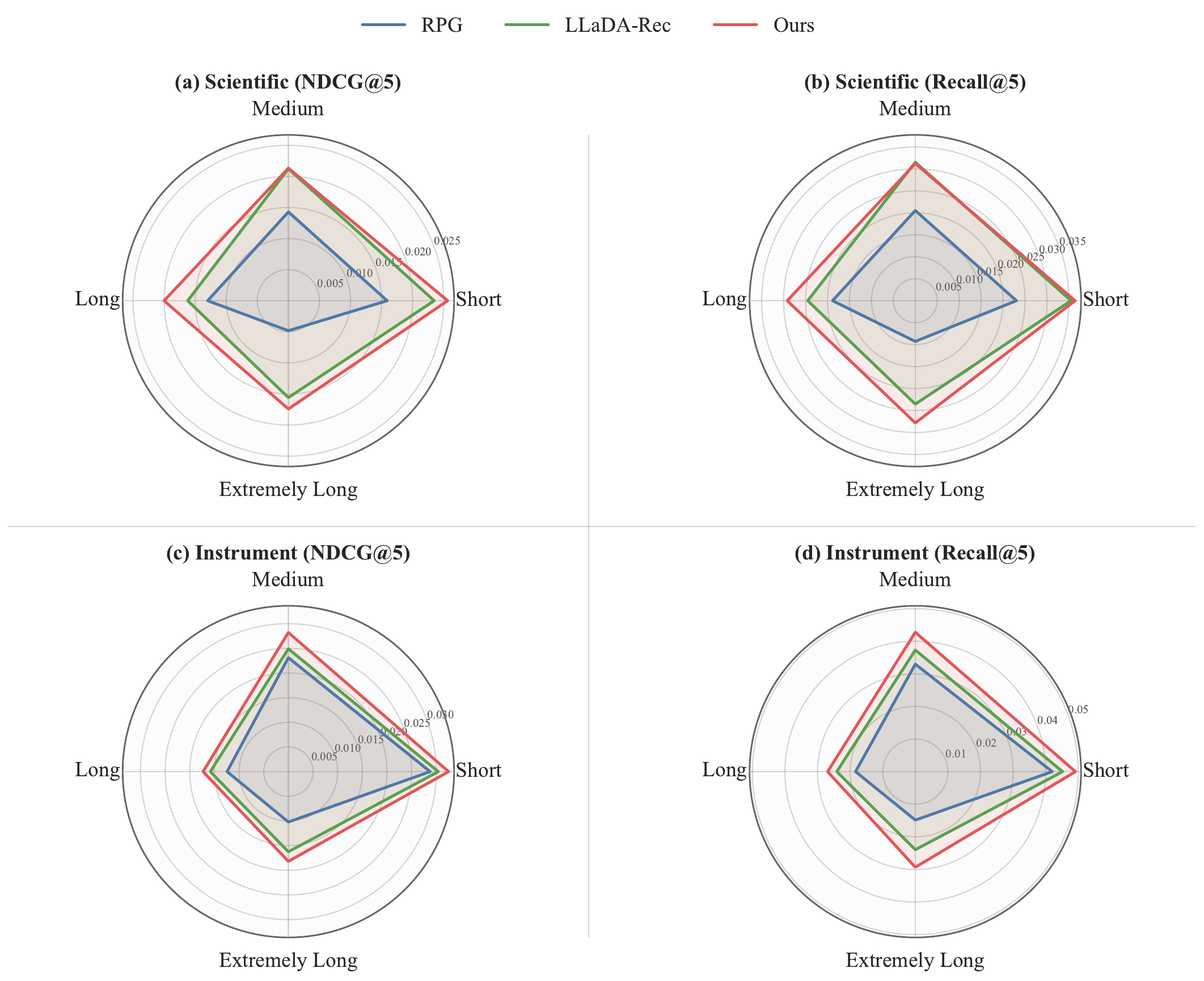}
    \caption{Performance comparison across different user interaction sequence length groups on the Scientific and Instrument dataset.}
    \label{fig:seq_length}
\end{figure}

As illustrated in Figure \ref{fig:seq_length}, \ourname~ consistently outperforms the baselines across all length groups. Notably, RPG experiences a performance drop on extremely long sequences ($>20$). This degradation could be partially attributed to the unidirectional bias, where the AR model over-emphasizes recent items while forgetting early context. In contrast, diffusion-based models exhibit better resilience due to their bidirectional attention. Furthermore, \ourname~ surpasses LLaDA-Rec on short sequences ($<5$). Interestingly, the short-sequence group tends to achieve relatively strong results across different models, suggesting that these users may have more concentrated and less noisy preferences rather than representing purely cold-start cases. This is reasonable for Amazon datasets, where the average sequence length is only around 8, making the $<5$ group still close to the dominant interaction range. 
\subsubsection{Performance across Item Popularity}
Popularity imbalance remains a fundamental challenge in recommendation systems, as user interactions are typically concentrated on a small subset of highly frequent items while many others suffer from limited exposure. To better understand how our method behaves under different popularity regimes, we conduct a fine-grained analysis on the Scientific and Instrument dataset by partitioning items according to their interaction frequency.

Following standard practices, we sort all items based on their interaction frequencies in the training set and divide them into three disjoint groups: Head (the top 20\% most popular items), Mid (the next 60\% items), and Tail (the bottom 20\% infrequent items). We then compute NDCG@5 and Recall@5 for interactions in the test set where the target item belongs to these respective groups. We compare \ourname~ with the strongest autoregressive model RPG, and the discrete diffusion model LLaDA-Rec.

\begin{figure}[htbp]
    \centering
    \includegraphics[width=0.8\linewidth]{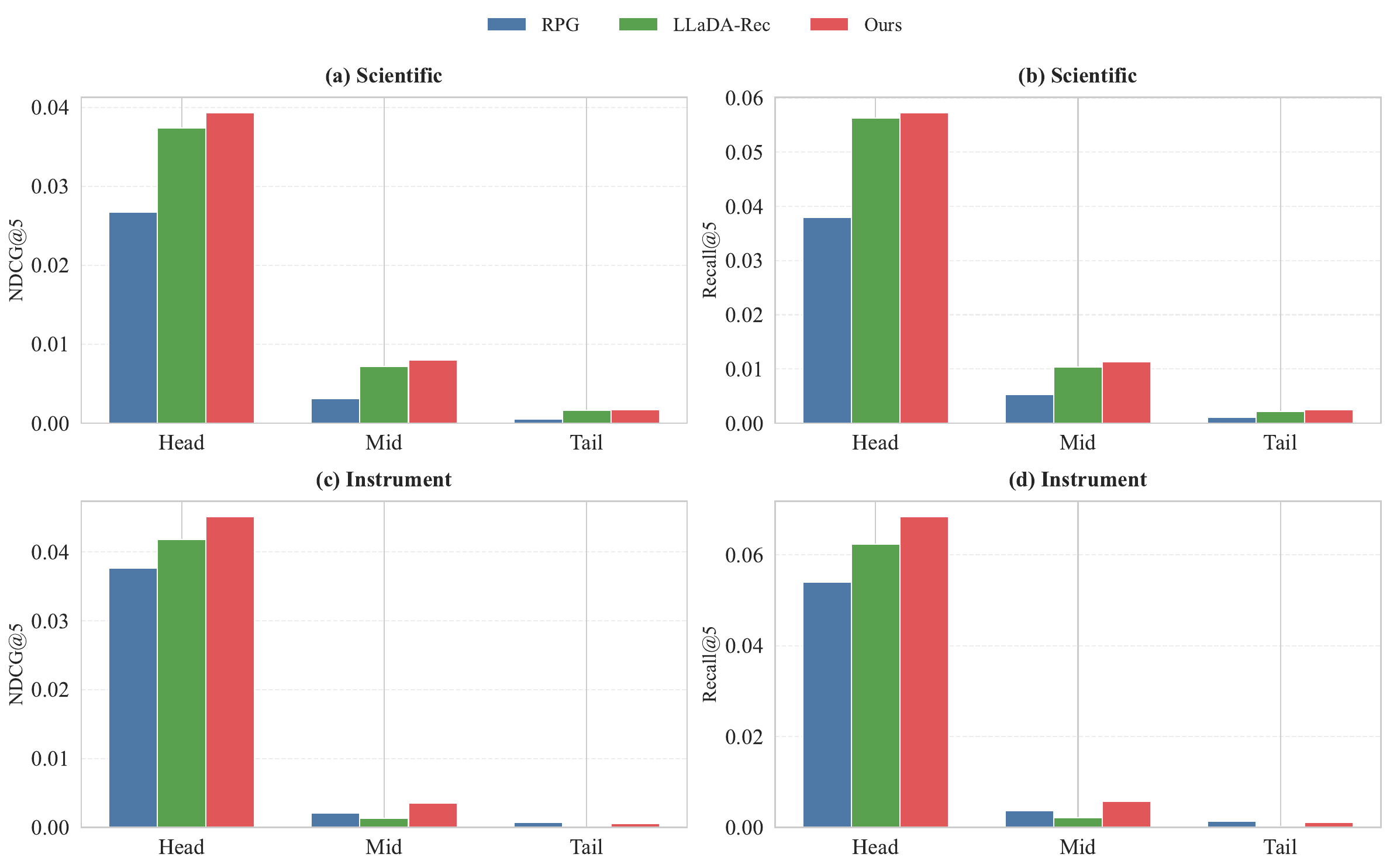}
    \caption{Performance comparison across Head, Mid, and Tail item groups on the Scientific and Instrument dataset.}
    \label{fig:itempop}
\end{figure}

As illustrated in Figure \ref{fig:itempop}, \ourname~ consistently outperforms LLaDA-Rec across head, mid, and tail item groups, with the clearest gains appearing on mid-frequency items. This suggests that our method is particularly effective at modeling items with moderate interaction frequency, where both collaborative dependency and transition diversity are important.
For tail items, the improvements are relatively smaller, since extremely sparse items still provide limited behavioral evidence for reliable generation. Nevertheless, on the Instruments dataset, \ourname~ still achieves better tail performance than LLaDA-Rec, indicating that our item co-occurrence-guided SID construction may help alleviate the representation isolation issue of infrequent items. While RPG maintains a marginal lead in the extreme tail of Instruments, this is likely due to its longer semantic IDs over-specializing to rare patterns. In contrast, \ourname~ prioritizes a more balanced representation, ensuring robust generalization across the entire popularity spectrum rather than over-fitting to the extreme tail.

\subsection{Ablation Study}
We conduct extensive ablation studies to analyze the contribution of each component in our framework, as summarized in Table~\ref{tab:ablation}.
We compare our model against the following variants: (1) \textit{w/o Schedule}, which replaces our item-based noise rescheduling strategy with a standard uniform noise schedule; (2) \textit{w/o II}, which removes the SVD-based collaborative features, relying solely on text semantics in SID generation; (3) \textit{w/o Schedule and II}, which simultaneously removes both the adaptive noise scheduling mechanism and the collaborative structural features used in SID construction; and (4) \textit{RQ-VAE}, which replaces our RQ-KMeans tokenizer with RQ-VAE. 
As shown in Table~\ref{tab:ablation}, we observe:
\begin{itemize}
\item  \textbf{Impact of item-based adaptive noise rescheduling:} \textit{w/o Schedule} leads to the most significant performance drop across all datasets (e.g., NDCG@1 drops by over 15\% on Scientific). This confirms that our noise rescheduling is the core engine of \ourname, enabling the model to adaptively learn item-based collaborative filtering signals that uniform noise scheduling cannot capture. 
\item   \textbf{Benefit of item co-occurrence-guided SID generation:} The performance decline in \textit{w/o II} highlights the necessity of fusing item co-occurrence information. While semantic features provide content understanding, the collaborative signals are crucial for capturing the structural behavior of users.
\item  \textbf{Superiority of RQ-KMeans:} Our RQ-KMeans tokenizer (\textit{Full}) consistently outperforms the RQ-VAE variants. This may be partly because clustering-based discretization avoids the instability that can occur when training neural tokenizers from scratch.
\item  \textbf{Synergy of two components:} The performance drops to its lowest point when both modules are removed (\textit{w/o Schedule and II}), indicating that the two contributions are mutually reinforcing rather than redundant.
\end{itemize}

\begin{table}
\centering
\caption{Ablation study results on three Amazon datasets. ``II'' denotes item co-occurrence information, and ``Schedule'' denotes the diffusion adaptive noise rescheduling strategy.}
\setlength{\tabcolsep}{5pt}
\resizebox{0.75\textwidth}{!}{%
\begin{tabular}{l|l|ccccc}
\hline
Dataset & Method 
& NDCG@1 & NDCG@5 & NDCG@10 
& Recall@5 & Recall@10 \\
\hline

\multirow{5}{*}{Scientific}

& RQ-VAE
& 0.0120 & 0.0221 & 0.0269
& 0.0320 & 0.0487 \\

& w/o II
& 0.0109 & 0.0217 & 0.0268
& 0.0314 & 0.0479 \\

& w/o Schedule
& 0.0099 & 0.0205 & 0.0259
& 0.0312 & 0.0477 \\

& w/o Schedule and II
& 0.0097 & 0.0203 & 0.0253
& 0.0310 & 0.0474 \\

& \textbf{Full}
& \textbf{0.0122} & \textbf{0.0226} & \textbf{0.0279}
& \textbf{0.0327} & \textbf{0.0494} \\

\hline

\multirow{5}{*}{Instrument}

& RQ-VAE
& 0.0128 & 0.0270 & 0.0341
& 0.0408 & 0.0626 \\

& w/o II
& 0.0130 & 0.0265 & 0.0339
& 0.0403 & 0.0627 \\

& w/o Schedule
& 0.0127 & 0.0259 & 0.0331
& 0.0401 & 0.0612 \\

& w/o Schedule and II
& 0.0127 & 0.0258 & 0.0329
& 0.0391 & 0.0610 \\

& \textbf{Full}
& \textbf{0.0135} & \textbf{0.0273} & \textbf{0.0343}
& \textbf{0.0410} & \textbf{0.0629} \\

\hline

\multirow{5}{*}{Game}

& RQ-VAE
& 0.0208 & 0.0426 & 0.0532
& 0.0641 & 0.0973 \\

& w/o II
& 0.0209 & 0.0423 & 0.0530
& 0.0633 & 0.0964 \\

& w/o Schedule
& 0.0206 & 0.0413 & 0.0523
& 0.0628 & 0.0957 \\

& w/o Schedule and II
& 0.0205 & 0.0409 & 0.0514
& 0.0610 & 0.0936 \\

& \textbf{Full}
& \textbf{0.0214} & \textbf{0.0434} & \textbf{0.0542}
& \textbf{0.0653} & \textbf{0.0987} \\

\hline
\end{tabular}
}

\label{tab:ablation}
\end{table}

\subsubsection{Effect of Noise Rescheduling}
\begin{figure}
    \centering
    \includegraphics[width=1\linewidth]{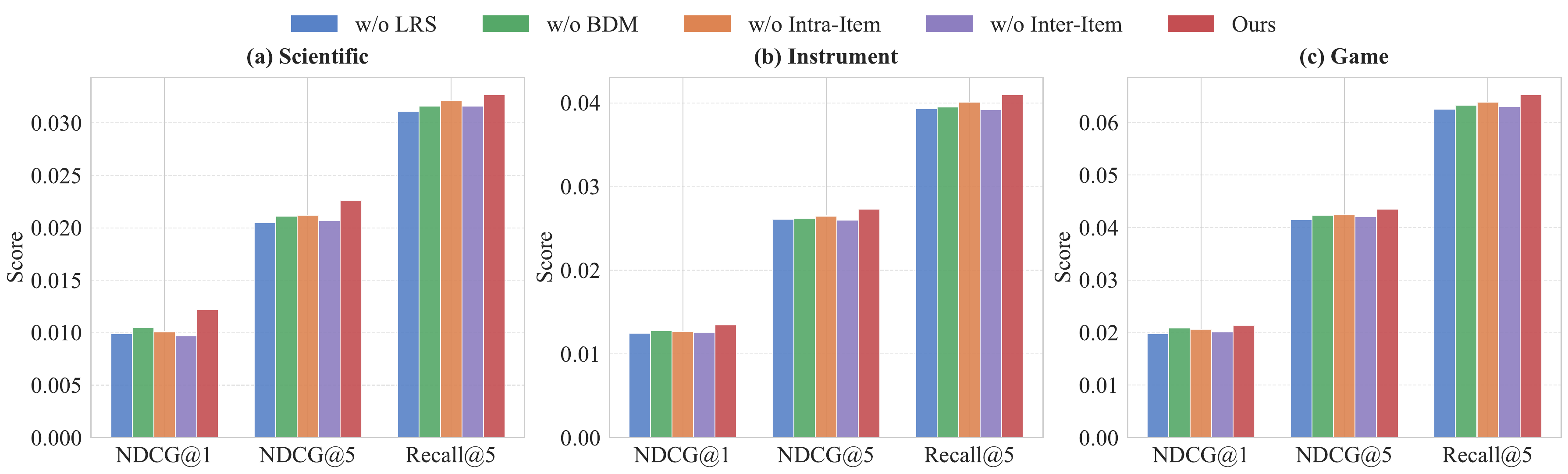}
    \caption{Ablation study on different noise rescheduling strategies. }
    \label{fig:ablation_resc}
\end{figure}

To further investigate the effectiveness of the proposed adaptive noise rescheduling mechanism, we compare \ourname~ with several variants in Figure~\ref{fig:ablation_resc}. The evaluated variants include: 
(1) \textit{w/o LRS}, which removes the geometric contextual recoverability modeling and only retains behavior-aware dependency modeling; 
(2) \textit{w/o BDM}, which removes the adaptive item-level dependency modeling and only preserves the local recoverability estimation; 
(3) \textit{w/o Inter-Item/Intra-Item}, which separately removes inter-item or intra-item contextual dependencies from the local recoverability estimation module.

From the results, removing \textit{local recoverability estimation} causes the most significant performance degradation, indicating that contextual recoverability serves as the primary source of denoising guidance in our adaptive rescheduling framework. By explicitly modeling the availability of surrounding unmasked contexts, the model can better distinguish easy-to-recover tokens from highly uncertain ones, thereby allocating denoising supervision more effectively. Without this mechanism, the diffusion process degenerates toward a nearly uniform denoising strategy, substantially weakening the ability to exploit structural collaborative signals.

Removing \textit{behavior-aware dependency modeling} also leads to a noticeable performance drop, although the degradation is smaller than removing local recoverability estimation. This suggests that adaptive item-level dependency modeling provides important complementary information beyond static structural priors. In particular, the behavior-aware dependency module enables the model to dynamically capture user-specific collaborative patterns, helping the denoising process focus on semantically correlated items within different behavioral contexts.

We further analyze the contributions of inter-item and intra-item dependencies within local recoverability estimation. Removing \textit{inter-item} contextual modeling results in a significantly larger decline than removing \textit{intra-item} dependencies, demonstrating that collaborative relationships across items constitute the dominant source of contextual recoverability in recommendation sequences. Since user behaviors are inherently driven by sequential item interactions, neighboring and correlated items provide strong structural cues for recovering masked tokens. In contrast, intra-item dependencies mainly refine the semantic consistency within each semantic ID, yielding a relatively smaller but still consistent contribution.

\subsubsection{Effect of Collaborative Signals} 
To evaluate how different forms of collaborative signal construction affect RQ-KMeans quantization, we compare several representative strategies, including SVD-based matrix factorization, Node2Vec~\cite{grover2016node2vec}, and a graph-based LightGCN~\cite{he2020lightgcn} encoder. 

\begin{figure}
    \centering
    \includegraphics[width=0.7\linewidth]{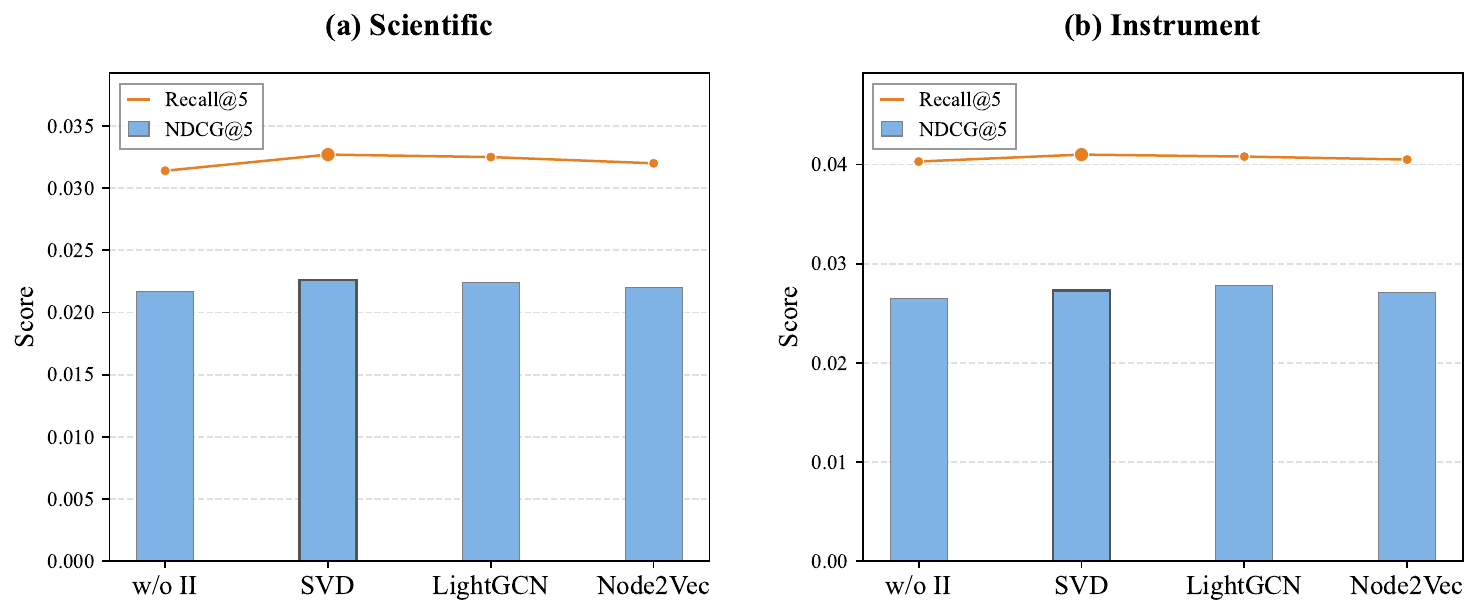}
    \caption{Ablation study on different forms of collaborative signal construction.}
    \label{fig:svdab}
\end{figure}

In Figure~\ref{fig:svdab}, when introducing collaborative signals, both SVD-based and LightGCN-based approaches achieve clear improvements over the \textit{w/o II} across datasets. However, their performance is largely comparable on both Scientific and Instrument, suggesting that first-order co-occurrence signals already provide a strong inductive bias for constructing SID representations, while higher-order propagation does not always translate into additional gains under our discrete diffusion framework.
In contrast, Node2Vec performs noticeably worse than both SVD and LightGCN. We attribute this to its reliance on homogeneous random walks, which tend to dilute fine-grained user-item interaction structure and are less aligned with the bipartite nature of recommendation data.
Overall, these results indicate that incorporating explicit collaborative structure is beneficial, while the specific choice between SVD and LightGCN has limited impact in our setting. This suggests that the effectiveness of SID construction is primarily driven by the presence of collaborative information itself rather than the complexity of the encoder used to extract it.

\subsection{Hyper-parameter Analysis}












We investigate the impact of two critical hyperparameters on model performance: the dimensionality of collaborative embeddings and the sharpness of the geometric distribution used in our noise rescheduling.

\subsubsection{Impact of Collaborative Embedding Dimension.} 
\begin{figure}
    \centering
    \includegraphics[width=0.7\linewidth]{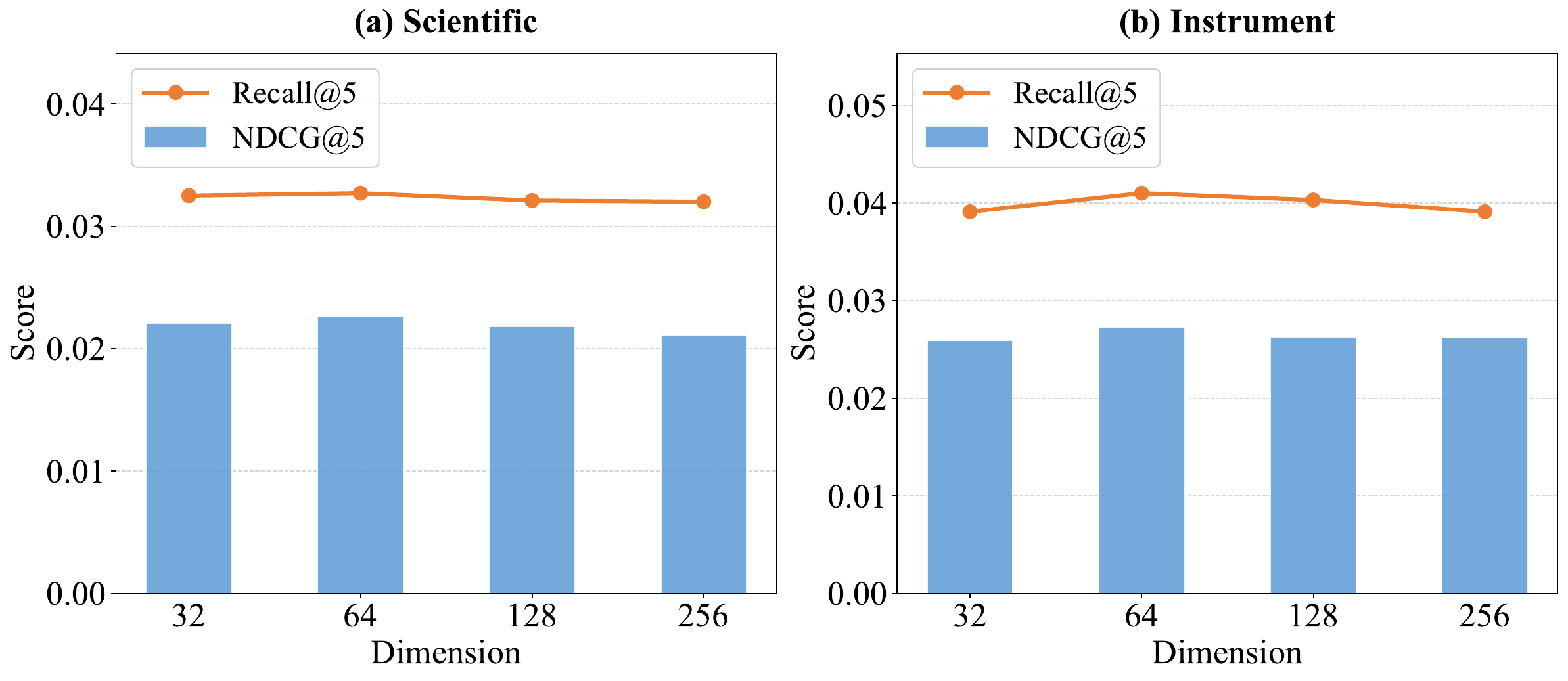}
    \caption{Hyper-parameter study with different dimensionalities for collaborative embedding.}
    \label{fig:hyper_ii}
\end{figure}

We first vary the dimension of the SVD-decomposed item co-occurrence features within $\{32, 64, 128, 256\}$. As illustrated in Figure~\ref{fig:hyper_ii}, the recommendation performance initially improves as the dimension increases from 32 to 64. This indicates that a sufficient embedding size is necessary to encode the complex structural dependencies between items. However, further increasing the dimension to 128 or 256 yields negligible gains or even slight performance degradation. This suggests that 64 dimensions offer an optimal balance between representation capacity and noise suppression, effectively capturing collaborative signals without overfitting to sparse co-occurrence patterns. Consequently, we set the dimension to 64.
\subsubsection{Impact of Sharpness in Noise Rescheduling.} 
\begin{figure}
    \centering
    \includegraphics[width=0.6\linewidth]{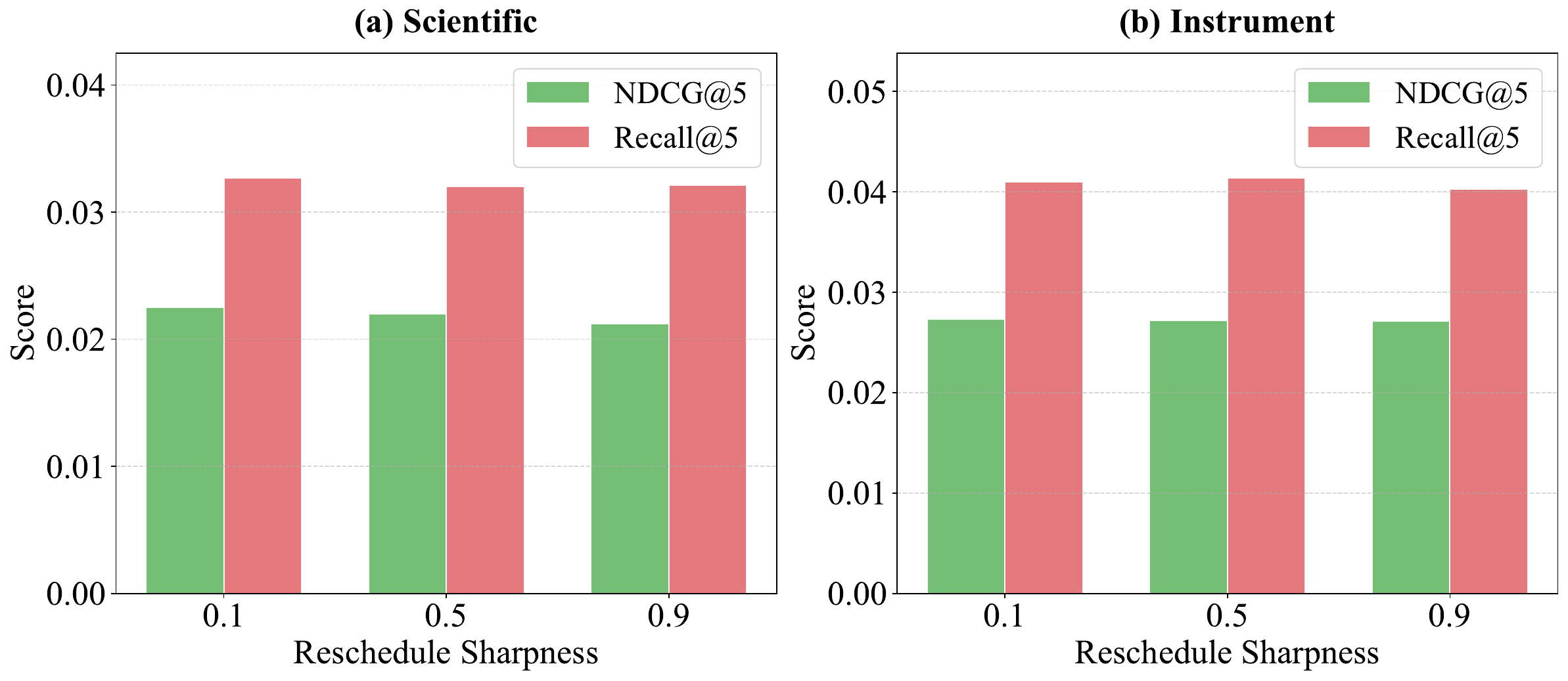}
    \caption{Hyper-parameter study of the sharpness in the geometric distribution.}
    \label{fig:hyper_reschedule}
\end{figure}

Additionally, we perform a sensitivity analysis on the sharpness parameter $p$ within the geometric distribution employed for our noise rescheduling, varying it across $\{0.1, 0.5, 0.9\}$. As shown in Figure~\ref{fig:hyper_reschedule}, the model exhibits robust performance across different settings. Based on these observations, we set $p=0.1$ for all datasets.


\subsection{Efficiency Analysis}

\subsubsection{Complexity Analysis.}
We briefly analyze the computational complexity of the proposed \ourname~, focusing on our primary contribution: the item-based adaptive noise rescheduling mechanism. For a user history of length $N$ with SIDs of length $L$, the total sequence length is $M = (N-1) \times L$. 

The standard bidirectional Transformer encoder incurs a self-attention complexity of $\mathcal{O}(M^2 d)$, where $d$ is the hidden state dimension. In our proposed rescheduling module, computing the structural geometric decay $w^{struct}_{ij}$ requires $\mathcal{O}(M^2)$ operations. Concurrently, extracting the attention scores $A_{ij}$ directly reuses the final hidden states from the Transformer, incurring $\mathcal{O}(M^2 d)$ complexity. Therefore, the overall time complexity of \ourname~ remains strictly bounded by $\mathcal{O}(M^2 d)$. It operates in the same asymptotic complexity class as standard Transformer-based baselines (e.g., BERT4Rec, LLaDA-Rec), guaranteeing that our fine-grained denoising guidance is achieved without compromising training scalability.

\subsubsection{Training Efficiency.}
We analyze the training cost from two aspects: SID generation and discrete diffusion model optimization.
As shown in Table~\ref{tab:train_efficiency}, RQ-VAE-based SID generation requires expensive pretraining. 
In contrast, our item co-occurrence-guided SID generation incurs only a fraction of the computational cost, including the construction of the co-occurrence matrix, its SVD decomposition, and the RQ-KMeans clustering.
For our model training, the proposed item-based adaptive noise rescheduling introduces marginal overhead (around 1\%) compared to LLaDA-Rec~\cite{shi2025lladarec}.

\begin{table}[t]
\caption{Training efficiency comparison in Scientific.}
\centering
\resizebox{0.6\textwidth}{!}{%
\begin{tabular}{clccc}
\toprule
\text{Stage} & \text{Method} & \text{Time/Epoch (s)} & \text{Total (h)} & \text{Cost} \\
\midrule
\multirow{2}{*}{\shortstack{Item\\Tokenization}} 
 & RQ-VAE & 0.648 & 1.799 & 1.000$\times$ \\
 & \text{Ours} & -- & \textbf{0.106} & \textbf{0.059}$\times$ \\
\midrule
\multirow{2}{*}{\shortstack{Generative\\Training}} 
 & LLaDA-Rec & 87.402 & 3.641 & 1.000$\times$ \\
 & \text{\ourname} & 88.351 & 3.681 & 1.011$\times$ \\
\bottomrule
\end{tabular}
}
\label{tab:train_efficiency}
\end{table}


\subsubsection{Inference Efficiency.}
We further analyze the efficiency of the proposed constrained diffusion inference. 
Although the decoding process maintains a candidate item set to ensure valid SID generation, the effective token vocabulary shrinks rapidly at each diffusion step. 
As shown in Table~\ref{tab:inference_efficiency}, We scale the beam size from 1 to 50 to evaluate the total inference time on the Scientific test set. 
While the total overhead for both models scales linearly with the beam size, their execution times remain nearly identical across all configurations. Our method achieves a comparable per-user latency to LLaDA-Rec with Beam=50 (10.17 ms vs. 10.06 ms). 
This close match indicates that \ourname~  mitigates potential computational burdens, introducing negligible inference overhead compared to LLaDA-Rec while achieving superior recommendation performance.
\begin{table}[htbp]
\centering
\caption{Inference efficiency comparison with different beam sizes in Scientific.}
\label{tab:inference_efficiency}
\small %
\begin{tabular}{llcccccc}
\toprule
\textbf{Method} & \textbf{Metric} & \textbf{Beam=1} & \textbf{Beam=10} & \textbf{Beam=20} & \textbf{Beam=30} & \textbf{Beam=40} & \textbf{Beam=50} \\
\midrule
\multirow{2}{*}{LLaDA-Rec} & Total Time (s) & 0m 13s & 1m 46s & 3m 22s & 5m 04s & 6m 46s & 8m 33s \\
& Time / User (ms) & 0.25 & 2.08 & 3.96 & 5.96 & 7.96 & 10.06 \\
\midrule
\multirow{2}{*}{~\ourname} & Total Time (s) & 0m 13s & 1m 46s & 3m 24s & 5m 06s & 6m 50s & 8m 39s \\
& Time / User (ms) & 0.25 & 2.08 & 4.00 & 6.00 & 8.04 & 10.18 \\
\bottomrule
\end{tabular}
\end{table}

\begin{table}
    \centering
    \caption{Case study on three sampled items. We compare the number of ground-truth co-occurring items found within the Top-20 nearest neighbors in both the embedding space and the SID space. ``Pre'' and ``Post'' denote before and after integrating the item co-occurrence features. }
    \label{tab:multi_case_study}
    \resizebox{1\textwidth}{!}{
    \begin{tabular}{c|l|ccc|l}
        \toprule
        \multirow{2}{*}{\text{Item}} & \multirow{2}{*}{\text{Metric Space}} & \multicolumn{3}{c|}{\text{Co-occur Hits@20}} & \multirow{2}{*}{\text{Co-occurring Neighbors (Post-integration)}} \\
        \cmidrule(lr){3-5}
         & & \text{Pre} & \text{Post} & \text{Gain ($\uparrow$)} & \\
        \midrule
        
        \multirow{2}{*}{\text{Item \#710}} 
         & Embedding & 3 & 5 & +2 & \{757, 732, 9894, \textbf{750, 1104}\} \\
         & SIDs & 1 & \text{8} & \text{+7} & \makecell[l]{\{\textbf{732, 719, 750, 1104, 755, 758, 751}, 757\}} \\
        \midrule
        
        \multirow{2}{*}{\text{Item \#6278}} 
         & Embedding & 5 & 9 & +4 & \makecell[l]{\{9430, 3671, 23939, 2229, \textbf{3021, 17400, 24692, 21922}, 17350\}} \\
         & SIDs & 4 & \text{12} & \text{+8} & \makecell[l]{\{3671, \textbf{3717}, \textbf{879}, 2371, \textbf{2250, 2230, 1613, 2268, 1169}, 998, 2229, \textbf{1774}\}} \\
        \midrule
        
        \multirow{2}{*}{\text{Item \#219}} 
         & Embedding & 4 & 8 & +4 & \{975, \textbf{971}, \textbf{20665, 977}, 214, 13070, 9644, \textbf{6499}\} \\
         & SIDs & 3 & \text{7} & \text{+4} & \{220, 214, \textbf{516, 6499, 13148}, 13070, \textbf{971}\} \\
         
        \bottomrule
    \end{tabular}
    }
\end{table}
\subsection{Case Studies}

To rigorously evaluate the effectiveness of incorporating item co-occurrence features in the SID generation, we examine three randomly sampled items (\#710, \#6278, and \#219) with distinct characteristics. For each item, we retrieve the Top-20 nearest neighbors based on continuous embedding similarity and discrete SIDs overlap, respectively. We then verify how many of these neighbors appear in the ground-truth global co-occurrence matrix.
As detailed in Table~\ref{tab:multi_case_study}, our method yields consistent improvements across all cases.
Embedding space refinement: after integrating co-occurrence features, the embedding space becomes more collaborative.
For instance, in Item \#710, the number of co-occurring neighbors in the top-20 list increases from 3 to 5.
Significant SID refinement: the most notable gains are observed in the SID space.
For instance, in Item \#6278, the number of co-occurring neighbors found by SID retrieval tripled (from 4 to 12) after feature integration.
This confirms that our quantizer, when guided by the co-occurrence feature, successfully encodes collaborative information into the SIDs.
Boldface indicates the gain in co-occurring hits.

\section{Conclusion}
This paper presents \ourname, a novel generative recommendation framework that integrates item-based collaborative structures into discrete diffusion models. The core of our contribution is the item-based adaptive noise rescheduling strategy, which provides structure-aware denoising guidance by adaptively adjusting the noise level according to intra-item and inter-item contexts. Complementary to this, we introduce an item co-occurrence-guided SID generation mechanism that injects item co-occurrence priors for diffusion training. Empirical results on various benchmarks confirm the effectiveness of our approach, demonstrating that aligning the denoising process with the collaborative signals in recommendation data significantly improves generative performance. In future work, we aim to further advance generative recommendation models by integrating multimodal information.
\bibliographystyle{ACM-Reference-Format}
\bibliography{sample-base}

\end{document}